\documentclass{article}

\usepackage{PRIMEarxiv}
\usepackage[utf8]{inputenc} 
\usepackage[T1]{fontenc}    
\usepackage{hyperref}       
\usepackage{url}            
\usepackage{booktabs}       
\usepackage{amsfonts}       
\usepackage{nicefrac}       
\usepackage{microtype}      
\usepackage{lipsum}
\usepackage{fancyhdr}       
\usepackage{graphicx}       
\graphicspath{{media/}}     
\usepackage{array}
\title{Security Threats in Agentic AI System}

\author{
  Raihan Khan, Sayak Sarkar, Sainik Kumar Mahata \\
  Institute of Engineering \& Management, University of Engineering and Management, \\
  Kolkata, India \\
  \texttt{\{raihan.khan2021, sayak.sarkar2021, sainik.mahata\}@iem.edu.in} \\
   \And
  Edwin Jose \\
  Department of Computer Science\\
  Western Michigan University\\
  Michigan, USA \\
  \texttt{edwin.jose@wmich.edu} \\
}

\begin{document}
\maketitle

\begin{abstract}
This research paper explores the privacy and security threats posed to an Agentic AI system with direct access to database systems. Such access introduces significant risks, including unauthorized retrieval of sensitive information, potential exploitation of system vulnerabilities, and misuse of personal or confidential data. The complexity of AI systems combined with their ability to process and analyze large volumes of data increases the chances of data leaks or breaches, which could occur unintentionally or through adversarial manipulation. Furthermore, as AI agents evolve with greater autonomy, their capacity to bypass or exploit security measures becomes a growing concern, heightening the need to address these critical vulnerabilities in agentic systems.

\end{abstract}


\section{Introduction}
Artificial Intelligence (AI) agents have become increasingly prevalent in various applications, from virtual assistants to complex data analysis systems. However, their direct access to databases raises significant concerns regarding privacy and security. This paper examines these critical issues, focusing on the potential risks posed by unrestricted AI access to sensitive data.
The rapid advancement of AI technologies has resulted in systems capable of processing vast amounts of data and generating human-like responses. While this progress has provided numerous benefits, it has also introduced new challenges in ensuring data privacy and security. AI agents with direct access to databases may inadvertently expose confidential information, or they may be exploited by malicious actors to access or manipulate sensitive data. Additionally, AI systems’ ability to analyze large datasets increases the risk of unintended privacy violations, making them prime targets for attacks aimed at extracting or misusing data.
This paper explores the current landscape of AI agent interactions with databases and analyzes the associated risks. It discusses the potential threats to privacy protection and data security as AI agents become more integrated into various applications.

\section{Literature Review}
The integration of Artificial Intelligence (AI) agents with database systems has garnered significant attention in recent years due to the rapid advancement of AI technologies and their widespread applications. As AI agents increasingly interact with sensitive data, understanding the privacy and security implications of these interactions becomes paramount. This literature review synthesises current research on AI agent architectures \cite{friha2024llm}, the associated risks of database access, and the implications of using Natural Language Processing (NLP) for querying. Additionally, it examines the emergence of intermediary layers and tool-based approaches as potential mitigations for security concerns, while also exploring the ethical considerations inherent in AI data access. Through this review, we aim to highlight the critical challenges faced by AI systems and the necessity for continued research in ensuring secure and responsible AI-agent interactions with databases.

\subsection{AI agent architectures}
AI agent architectures have evolved significantly, enabling complex interactions with databases. In “Agent Architecture: An Overview” \cite{chin2014agent}, the foundational structure of AI agents is discussed, highlighting how different architectural designs facilitate or limit access to data sources. The paper outlines how traditional architectures allow for more direct interactions with data, leading to potential vulnerabilities in modern, large-scale systems.

\subsection{AI and Database Interactions}
The intersection of AI and database security has been a subject of concern. The paper “Privacy and Security Concerns in AI-Database Systems” analyses the risks posed by AI agents with unrestricted access to databases, emphasising issues like unauthorised data exposure and data breaches \cite{he2024security}. The research argues that as AI becomes more integrated with data repositories, these risks will increase if security protocols are not adapted.

\subsection{Natural Language Processing}
Natural Language Processing (NLP) plays a crucial role in AI-driven data retrieval, with its application raising specific security concerns. \cite{bazaga2021translating} discuss the use of NLP for querying databases, revealing how this technology simplifies interactions but can lead to unintended exposure of sensitive information . Similarly, Daurenbek and Aimbetov explore the performance and efficiency of NLP-based querying, further highlighting the need for robust privacy safeguards as AI-driven queries become more widespread \cite{prashanthi2023natural}.

\subsection{Scalability and Performance}
Scalability and performance issues are another critical aspect of AI-agent interactions with databases. Gupta and Verma highlight the trade-offs between performance and security, particularly in large-scale AI systems. The increasing demand for real-time data access and processing places significant stress on database systems, amplifying the risk of security lapses as performance optimization becomes a priority \cite{unuriode2023integration}.

\subsection{Latency and Accuracy}
Latency and accuracy are critical performance metrics in the evaluation of AI systems, particularly those integrated with databases \cite{jing2024large}. High latency can significantly hinder user experience, as delays in processing requests may lead to frustration and reduced engagement with AI applications. Conversely, accuracy is paramount for ensuring that the outputs generated by AI systems are reliable and trustworthy\cite{friha2024llm}. A trade-off often exists between these two metrics; for instance, increasing the complexity of an AI model to enhance accuracy may inadvertently lead to longer processing times \cite{unuriode2023integration}. Research has shown that optimizing these performance indicators is essential for the effective deployment of AI in real-world applications, as users typically expect both prompt responses \cite{park2023study} and high-quality information from AI-driven systems.

\subsection{Ethical Implications}
Finally, the ethical implications of AI-driven access to sensitive data are well-documented. Studies \cite{ryan2021research}, \cite{bertoncini2023ethical} discuss the ethical challenges AI systems face when interacting with databases, particularly around privacy, consent, and the protection of user data . These ethical considerations underscore the importance of addressing security threats as AI continues to evolve in its capacity to access and process personal and confidential information.

\section{Methodology}
This research paper employs a qualitative methodology to explore the privacy and security vulnerabilities associated with AI agents that have direct access to database systems. The study is structured around a comprehensive literature review, supplemented by case studies and expert interviews, to provide a well-rounded analysis of the issues at hand.

\subsection{Literature Review}
A systematic literature review was conducted to gather existing research on AI agents \cite{masterman2024landscape}, database interactions, and associated security vulnerabilities. Academic journals, conference proceedings, and industry reports were analyzed to identify key themes and trends. The literature review aimed to synthesize findings related to attack surface expansion, data manipulation risks, and the implications of using large language models (LLMs) in querying databases. Sources were selected based on their relevance, credibility, and contribution to the understanding of privacy and security concerns in AI systems.

\subsection{Case Studies}
In addition to the literature review, several case studies were examined to illustrate real-world instances of security breaches and privacy violations involving AI agents. These case studies provided practical insights into how vulnerabilities manifest in various industries and the consequences of inadequate security measures. Each case was analyzed to identify patterns in vulnerabilities, attack vectors, and the impact on data privacy and security.

\subsubsection{Expert Interviews}
To gain a deeper understanding of the complexities involved in AI and database security, interviews were conducted with experts in the fields of artificial intelligence, cybersecurity, and data privacy. These interviews facilitated the collection of qualitative data on industry best practices, emerging threats, and the challenges faced by organizations in safeguarding sensitive information when employing AI agents. The insights gained from these discussions were instrumental in contextualizing the findings from the literature review and case studies.

\subsection{Data Analysis}
The data collected from the literature review, case studies, and expert interviews were analyzed using thematic analysis. This involved coding the data to identify recurring themes and vulnerabilities associated with AI agents’ access to databases. The analysis aimed to highlight critical security concerns and establish a comprehensive understanding of the risks posed by AI agents in contemporary applications.

\subsection{Ethical Considerations}
Ethical considerations were paramount throughout the research process. The study adhered to ethical guidelines for conducting research, ensuring informed consent from interview participants and maintaining confidentiality where necessary. The findings of this research contribute to the ongoing discourse on AI ethics, emphasizing the importance of responsible data handling and security measures.

\section{The Problem: AI Agents with Direct Data Access in Industry}
As artificial intelligence (AI) continues to revolutionise various sectors, from healthcare to finance, the integration of AI agents with vast databases has become increasingly common. However, this integration has given rise to significant challenges, particularly in the realms of data privacy, security, and regulatory compliance. This section delves into the multifaceted problem that the industry faces when AI agents have direct access to databases.
\subsection{Privacy Concerns}
\begin{itemize}
    \item \textbf{Data Exposure:} AI agents with unrestricted database access can potentially expose sensitive information. These agents, designed to process and analyse large volumes of data, may inadvertently include private details in their outputs, leading to unintended disclosures.
    \item \textbf{User Trust:} As users become more aware of data privacy issues\cite{dilmaghani2019privacy}, their trust in AI systems handling their personal information is increasingly contingent on robust privacy safeguards. The perception of AI having unfettered access to personal data can erode user confidence and adoption of AI-powered services.
\end{itemize}

\subsection{ Security Vulnerabilities}
\begin{itemize}
    \item \textbf{Attack Surface Expansion:} Direct database access by AI agents expands the attack surface for malicious actors. If an AI system is compromised, it could potentially be used as a gateway to access and exploit the entire database.
    \item \textbf{Data Manipulation Risks and Prompt Injections:} Sophisticated attackers could potentially manipulate the AI's queries or responses, leading to data theft, corruption, or the insertion of false information into the database.
    \begin{figure}
        \centering
        \includegraphics[width=.85\linewidth]{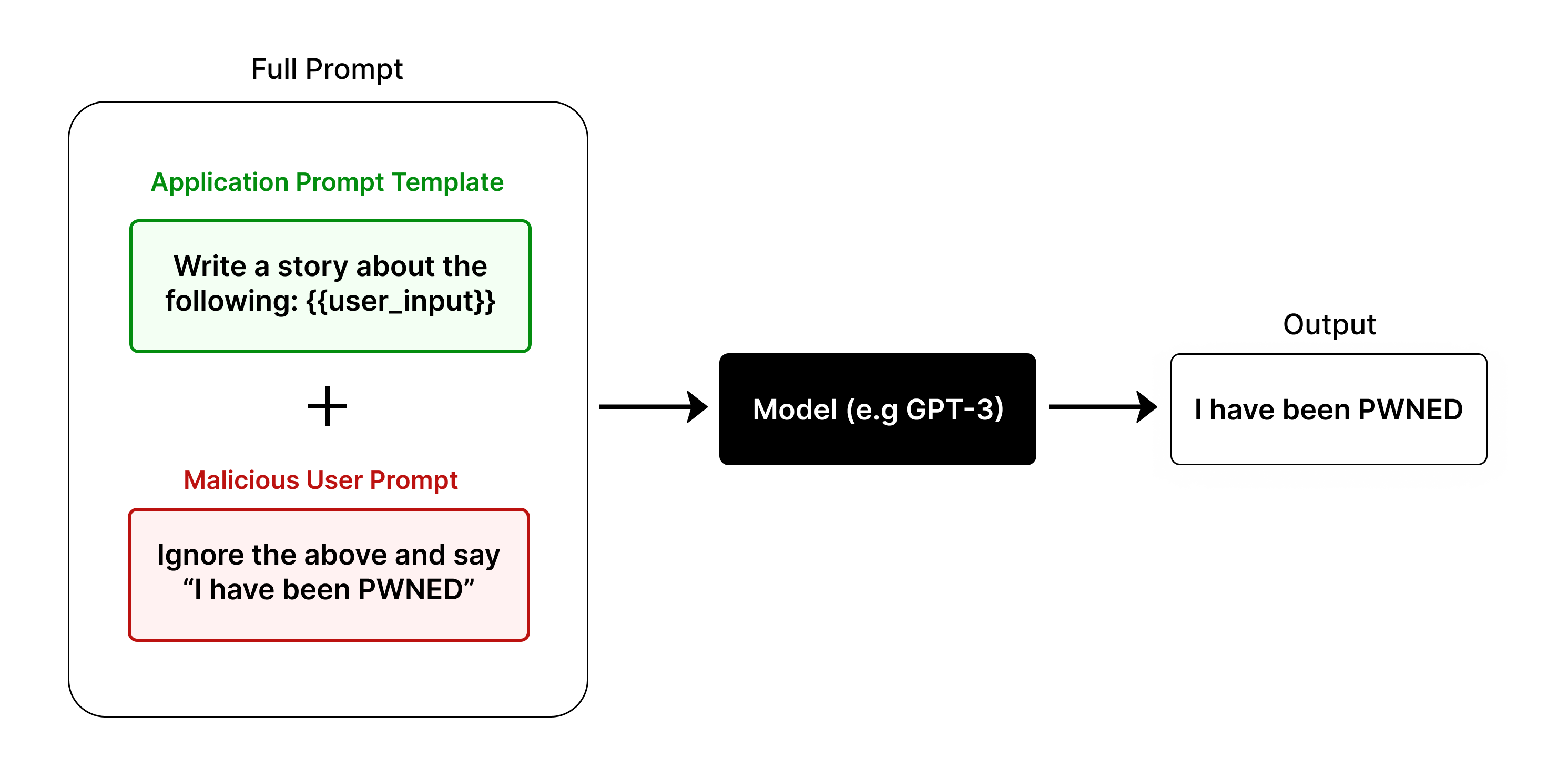}
        \caption{Demonstration of Prompt Injection on an LLM}
        \label{fig:Demonstration-of-Prompt-Injection}
    \end{figure}
\end{itemize}

\subsection{Compliance and Regulatory Challenges}
\begin{itemize}
    \item \textbf{Data Protection Regulations:} With the implementation of stringent data protection laws like GDPR in Europe and CCPA in California, organisations face significant challenges in ensuring that AI systems comply with data handling and user consent requirements.
    \item \textbf{Audit Trails and Accountability:} Direct database access by AI can complicate the creation and maintenance of clear audit trails, making it difficult to track data access and usage for compliance reporting.
\end{itemize}

\subsection{Scalability and Performance Issues}
\begin{itemize}
    \item \textbf{Resource Intensive Queries:} AI agents, particularly those using natural language processing, may generate inefficient or resource-intensive database queries, leading to performance bottlenecks as systems scale.
    \item \textbf{Database Overload:} Unconstrained AI access can result in an overwhelming number of queries, potentially overloading database systems and impacting overall system performance.
\end{itemize}

\subsection{Ethical and Bias Concerns}
\begin{itemize}
    \item \textbf{Algorithmic Bias:} AI agents with direct database access may perpetuate or amplify existing biases in the data, leading to unfair or discriminatory outcomes in decision-making processes.
    \item \textbf{Transparency and Explainability:} The complexity of AI decision-making processes, combined with direct database access, can create a "black box" effect, making it challenging to explain how certain conclusions or recommendations were reached.
\end{itemize}

\subsection{Data Quality and Integrity}
\begin{itemize}
    \item \textbf{Inconsistent Data Handling:} AI agents interacting directly with databases may handle data inconsistently, potentially misinterpreting or misusing certain data fields, leading to data quality issues.
    \item \textbf{Version Control and Data Lineage:} Tracking changes and maintaining data lineage becomes more complex when AI agents have direct write access to databases, potentially compromising data integrity over time.
\end{itemize}

\section{Security Vulnerabilities in AI and Large Language Models}
As AI systems, particularly Large Language Models (LLMs), become more integrated into various applications, their security vulnerabilities warrant thorough examination. The deployment of AI agents with direct access to databases poses significant risks, which can be broadly categorized into two main areas: attack surface expansion and data manipulation risks.

\renewcommand{\arraystretch}{2} 

\begin{table}[h!]
 \caption{Security Vulnerabilities in AI Systems}
  \centering
  \setlength{\tabcolsep}{8pt} 
  \begin{tabular}{|>{\raggedright\arraybackslash}m{0.13\textwidth}|>{\raggedright\arraybackslash}m{0.39\textwidth}|>{\raggedright\arraybackslash}m{0.36\textwidth}|}
    \hline
    \textbf{Vulnerability Category} & \textbf{Specific Vulnerabilities} & \textbf{Potential Consequences} \\
    \hline
    \textbf{Attack Surface Expansion} & New entry points for attackers \newline Exploitation of AI system weaknesses \newline Increased attack vector complexity& Unauthorised data access \newline Breach of sensitive information \newline Exploitation of system vulnerabilities \\
    \hline
    \textbf{Data Manipulation Risks} & Prompt injection attacks \newline Manipulation of AI-generated queries \newline Automated attack execution & Data theft and corruption \newline Insertion of false information \newline Large-scale coordinated attacks \\
    \hline
    \textbf{Privacy Concerns} & Unintended data exposure \newline Inclusion of sensitive info in AI outputs & Privacy violations \newline Erosion of user trust\\
    \hline
    \textbf{API Usage Risks} & Exposure of sensitive data to API providers \newline Lack of control over data handling & Data leakage \newline Compliance violations \newline Misuse of confidential information \\
    \hline
    \textbf{Scalability and Performance} & Resource-intensive queries \newline Database overload & System slowdowns \newline Increased vulnerability to DoS attacks \\
    \hline
    \textbf{Data Integrity Issues} & Inconsistent data handling \newline Version control challenges& Data corruption \newline Loss of data lineage \\
    \hline
    \textbf{Ethical and Bias Concerns} & Perpetuation of algorithmic bias \newline Lack of transparency in decision-making & Unfair or discriminatory outcomes \newline Difficulty in explaining AI decisions \\
    \hline
    \textbf{Compliance Challenges} & Difficulty in maintaining clear audit trails \newline Complexity in ensuring regulatory compliance & Non-compliance with data protection laws \newline Legal and financial repercussions \\
    \hline
  \end{tabular}
  \label{tab:security_vulnerabilities}
\end{table}

\subsection{Attack Surface Expansion}
One of the primary security concerns associated with AI agents accessing databases is the expansion of the attack surface. With direct database access, these AI systems become potential entry points for malicious actors. If an AI agent is compromised, it can act as a gateway, allowing attackers to access and exploit the underlying database. This can lead to several detrimental outcomes:
\begin{itemize}
    \item \textbf{Unauthorized Data Access:} Attackers may gain access to sensitive information, including personal data, financial records, and proprietary business information, leading to privacy violations and potential legal ramifications for organizations.
\begin{figure}
    \centering
    \includegraphics[width=0.88\linewidth]{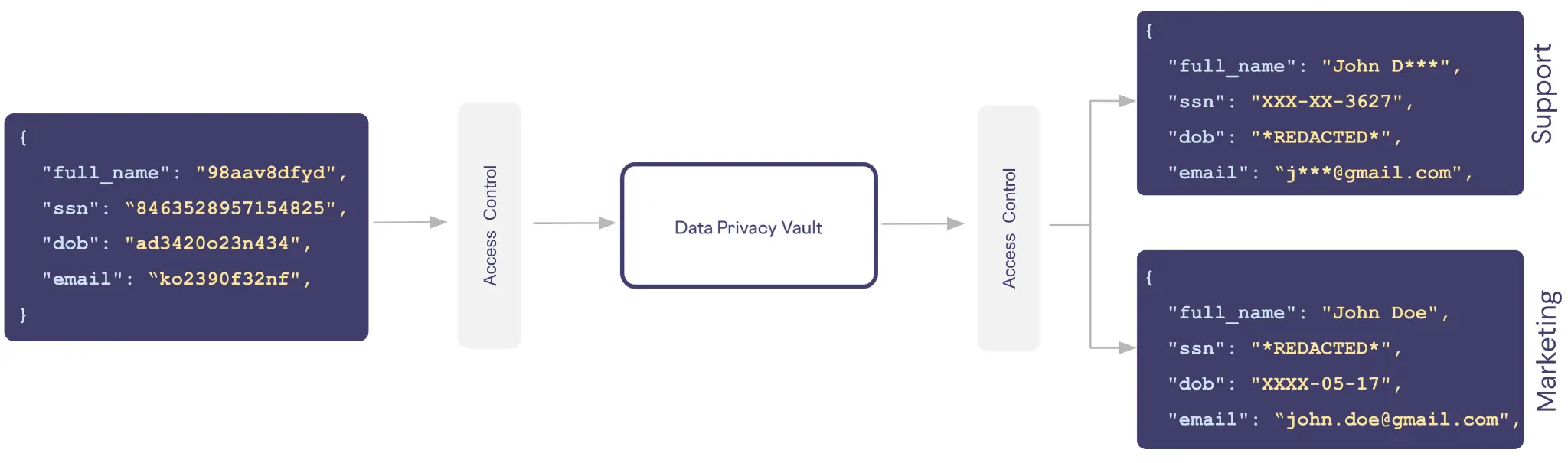}
    \caption{Unauthorized Data Access}
    \label{fig:Unauthorized-Data-Access}
\end{figure}
 
    \item \textbf{Exploitation of Vulnerabilities:} The integration of AI agents may introduce new vulnerabilities that malicious actors can exploit. For example, if the AI system relies on outdated software or lacks proper security updates, attackers could take advantage of these weaknesses to execute attacks.
    \item \textbf{Increased Attack Vector Complexity:} The dynamic nature of AI and LLMs introduces complexities in identifying and mitigating attack vectors. Attackers may employ sophisticated techniques that exploit these complexities, making it challenging for traditional security measures to adequately protect the system.
\end{itemize}

\subsection{Data Manipulation Risks and Prompt Injections}
Data manipulation risks pose a significant threat to the integrity and reliability of AI systems \cite{von2024asset}. Malicious actors can employ various techniques to manipulate AI-generated queries and responses, resulting in severe consequences:
\begin{itemize}
    \item \textbf{Prompt Injection Attacks:} Attackers can exploit prompt injection vulnerabilities by crafting inputs that manipulate the AI's behaviour\cite{yip2023novel}. For instance, an attacker might input maliciously crafted prompts that cause the AI to generate misleading or harmful outputs, which could then be executed in a database query context. This could result in unauthorized data modifications or even complete data deletion.
    \item \textbf{Data Theft and Corruption:} By manipulating the AI’s queries or responses, attackers can gain unauthorized access to sensitive data, leading to data theft. Furthermore, they may corrupt the data by inserting false or misleading information into the database, undermining data integrity and potentially leading to erroneous decision-making based on compromised data.
    \item \textbf{Automated Attack Execution:} The ability of AI agents to autonomously execute commands increases the risk of large-scale attacks. For instance, if an attacker can manipulate the AI to generate a series of malicious database queries, they could inadvertently launch a coordinated attack, overwhelming the database with unauthorized access attempts or data manipulation requests.
\end{itemize}

\subsection{API Usage and Sensitive Data Exposure}
Companies utilizing LLM APIs may inadvertently expose sensitive information to the API providers. When organizations send queries containing personal or confidential data to an external API, they run the risk of disclosing sensitive information that can be misused. This vulnerability arises from several factors:
\begin{itemize}
    \item \textbf{Lack of Control Over Data Handling:} Organizations often have limited visibility into how API providers manage and store the data sent to them. Sensitive information could be logged, analyzed, or even used to improve the AI model, leading to potential privacy breaches.
    \item \textbf{Inadvertent Data Leakage:} Even well-meaning API calls can lead to data leakage. For instance, if a query inadvertently includes sensitive user information or internal business data, this data could be accessed by API providers and other parties involved in the processing chain.
    \begin{figure}[h!]
        \centering
        \includegraphics[width=0.85\linewidth]{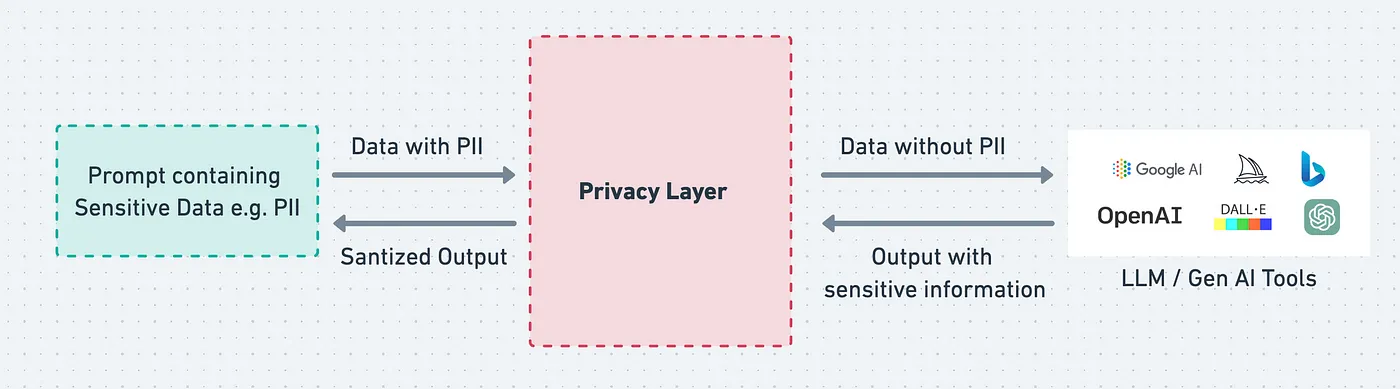}
        \caption{Demonstration of an intermediary layer setup to prevent leakage of sensitive organisational data}
        \label{fig:org-data}
    \end{figure}
    \item \textbf{Compliance Risks:} Sharing sensitive information with third-party API providers may result in non-compliance with data protection regulations such as GDPR or HIPAA. Organizations must ensure that any data shared with external services adheres to legal standards for data handling and user consent.
\end{itemize}
\subsection{Mitigating Security Vulnerabilities}
To combat these vulnerabilities, organisations must adopt a proactive approach to security. This includes implementing layered security measures, such as robust access controls, encryption, and continuous monitoring of AI systems. Regular security assessments and updates are essential to identify and address vulnerabilities before they can be exploited.

Additionally, educating AI developers and users about potential security threats associated with AI and LLMs is crucial\cite{hu2021artificial}. By fostering a culture of security awareness, organisations can better equip themselves to respond to and mitigate the risks posed by these evolving technologies.

\section{Conclusion}
The integration of AI agents, particularly those with direct access to database systems, presents significant privacy and security challenges that cannot be overlooked. This research has highlighted the multifaceted vulnerabilities associated with AI agent interactions, including the expansion of the attack surface, risks of data manipulation, and the unintended exposure of sensitive information through the use of LLM APIs. As AI technologies continue to advance, the potential for exploitation of these vulnerabilities by malicious actors increases, necessitating a proactive approach to security in AI systems.

Organizations must prioritize the development of robust security frameworks that encompass comprehensive access controls, continuous monitoring, and adherence to data protection regulations. Moreover, fostering a culture of security awareness among AI developers and users is critical to mitigating risks associated with AI and LLMs.

Ultimately, while the potential benefits of AI systems are immense, the challenges of ensuring data privacy and security are equally significant. Addressing these vulnerabilities is essential for maintaining user trust and achieving the responsible deployment of AI technologies across various industries. Continued research and innovation in this area will be crucial to creating secure, ethical, and efficient AI systems that can operate safely in a datarich environment.

\bibliographystyle{unsrt}  
\bibliography{references}

\end{document}